\documentstyle[12pt]{article}

\newcommand{\mybibitem}[1]{\bibitem{#1}}

\newcommand{\be}[1]{\begin{eqnarray} \label{#1} }
\newcommand{\eq}{\end{eqnarray}}

\def\a{\alpha}
\def\b{\beta}

\def\d{\delta}
\def\e{\epsilon}	
\def\f{\phi}	
\def\vf{\varphi}

\def\g{\gamma}

\def\p{\pi}			
\def\q{\theta}			

\def\D{\Delta}

\def\G{\Gamma}

\def\L{\Lambda}

\def\cf{\cal{F}}

\def\ci{\cal{I}}

\def\half{\frac{1}{2}}

\newcommand{\NPB}[1]{Nucl.\ Phys.\ {\bf B#1}}
\newcommand{\PLB}[1]{Phys.\ Lett.\ {\bf B#1}}

\newcommand{\PRD}[1]{Phys.\ Rev.\ {\bf D#1}}

\makeatletter
\@addtoreset{equation}{section}
\@addtoreset{equation}{subsection}
\def\theequation{\ifnum\value{section}=0 \arabic{equation}\ignorespaces
\else \ifnum\value{subsection}=0 \thesection.\arabic{equation}\ignorespaces
\else \thesection.\arabic{subsection}.\arabic{equation}\ignorespaces
                       \fi
                 \fi}
\@addtoreset{table}{section}
\@addtoreset{table}{subsection}
\def\thetable{\ifnum\value{section}=0 \arabic{table}\ignorespaces
\else \ifnum\value{subsection}=0 \thesection.\arabic{table}\ignorespaces
\else \thesection.\arabic{subsection}.\arabic{table}\ignorespaces
                    \fi
              \fi}
\makeatother

\newcommand{\bw}{{\bar W}}
\newcommand{\qb}{{\bar Q}}
\newcommand{\fb}{\bar {\phi}}

\newcommand{\FF}{{\cal F}}
\newcommand{\HH}{{\cal H}}

\newcommand{\ad}{{\dot\alpha}}
\newcommand{\bd}{{\dot\beta}}

\newcommand{\Del}{\nabla}
\newcommand{\Delb}{{\bar\nabla}}
\newcommand{\DD}{{\hbox{D}'}}
\newcommand{\val}[1]{\left<{#1}\right>}

\newcommand{\shalf}{\hbox{$\frac12$}} 
\newcommand{\squart}{\hbox{$\frac14$}} 
\newcommand{\too}{~&\to&~} 
\newcommand{\ret}{\nonumber\\}      
\begin{document}

\begin{titlepage}

\begin{flushright}
ITP-SB-96-34 \\
USITP-96-08\\
hep-th/9607089 \\
\end{flushright}

\begin{center}
\vskip3em
{\large\bf On $N=2$ low energy effective actions}

\vskip3em
{\normalsize U.\ Lindstr\"om\footnote{E-mail:ul@vanosf.physto.se}\\
\vskip .5em{\it
Institute of Theoretical Physics \\
University of Stockholm \\
Box 6730 \\
S-113 85 Stockholm SWEDEN}\\
\vskip2em

F.\ Gonzalez-Rey,\footnote{E-mail:glezrey@insti.physics.sunysb.edu}
M.\ Ro\v cek,\footnote{E-mail:rocek@insti.physics.sunysb.edu} and R.\ von
Unge\footnote{E-mail:unge@insti.physics.sunysb.edu}${}^,$\footnote{On leave from
Department of Physics, Stockholm University, Sweden}\\
\vskip .5em
{\it Institute for Theoretical Physics\\ State University of New York\\
 Stony Brook, NY 11794-3840, USA\\}
}
\end{center}

\vfill

\begin{abstract}

\noindent
We propose a Wilsonian action compatible with special geometry and higher
dimension $N=2$ corrections, and show that the holomorphic contribution $\FF$ to
the low energy effective action is independent of the infrared cutoff. We
further show that for asymptotically free $SU(2)$ super Yang-Mills theories, the
infrared cutoff can be tuned to cancel leading corrections to $\FF$. We also
classify all local higher-dimensional contributions to the $N=2$ superspace
effective action that produce corrections to the K\"ahler potential when reduced
to $N=1$ superspace.
\end{abstract}

\vfill
\end{titlepage}
\section{Introduction}\label{intro}

Recently, we computed the one-loop effective K\"ahler potential for $N=2$
supersymmetric QCD \cite{BMM}. For the 1-PI generating functional, we found
terms that violated special geometry; we were able to interpret them as
arising from higher dimension terms in $N=2$ superspace. Here we examine the
effects of these terms, and show that to quadratic order in fluctatuations
about the vacuum, they rescale the kinetic terms of the massive electrically
charged fields without changing their masses. For gauge group $SU(2)$, this
rescaling arises as a shift by a scheme dependent constant
independent of the moduli.

In \cite{BMM}, we did not consider whether even higher dimensional terms in
$N=2$ superspace could give rise to corrections to the K\"ahler potential. Here
we find all possible local higher dimension terms that can give such
contributions.

We also found that a naive attempt to define a Wilsonian effective
K\"ahler potential was incompatible with special geometry even for the massless
sector, and hence, with $N=2$ supersymmetry \cite{BMM}. Here we show that
some of the one-loop corrections to the effective K\"ahler potential give
rise to infrared divergent terms in the component expansion of the 1-PI
effective action, so we cannot ignore the problem. We propose a Wilsonian action
with a field dependent cutoff (somewhat in the spirit of
\cite{ds}) that is compatible with special geometry. We find that up to
possible rescalings of the ultraviolet cutoff, the holomorphic function $\cf$
\cite{calf} is not modified. We also find that for asymptotically free $SU(2)$
gauge theories, the constant shift in the coefficient of the charged field
kinetic terms can be tuned to zero, preserving in detail all the results of
\cite{sw}.

\section{Preliminaries}
We begin with a brief review of $N=2$ superspace (see \cite{BMM,book} for more
details): the chiral spinor coordinates $\{\q^{a\a}\}=\{\q^{1\a},\q^{2\a}\}$
transform as a doublet under a rigid $SU(2)$ group unrelated to the gauge
group, as do their complex conjugates $\{\bar\q^\ad_a\}$, and the spinor
derivatives corresponding to them. The rigid $SU(2)$ indices are
raised and lowered by the antisymmetric invariant tensor $C_{ab}$, with $C_{12}=
C^{12}=1$. Super Yang-Mills theory is described by gauge-covariant spinor
derivatives satisfying the constraints
\be{cons}
\{\Del_{a\a},\Del_{b\b}\}&=&iC_{ab}C_{\a\b}\,\bar W\ ,\ret
\{\Delb^a_\ad,\Delb^b_\bd\}&=&iC^{ab}C_{\ad\bd}\,W\ ,\ret
\{\Del_{a\a},\Delb^b_\bd\}&=&i\d_a^b\, \Del_{\a\bd}\ ,
\eq
where $W,\bw$ are chiral (antichiral) scalar superfield strengths
$\Delb^a_\ad W=\Del_{a\a}\bw=0$. The Bianchi identities imply the important
relation
\be{relb}
\Del_a^\a\Del_{b\a}W=C_{ac}C_{bd}\,\Delb^{d\ad}\Delb^c_\ad\bw\ .
\eq

The procedure for reducing $N=2$ superfields and action to $N=1$ form is
well known: one defines $N=1$ superfield components as
\be{compo}
\f\equiv W|\ ,\ \ W_\a\equiv -\Del_{2\a}W|\ ,
\eq
where the bar denotes setting $\q^{2\a}=\bar\q_2^\ad=0$. One
identifies $\Del_1$ with the $N=1$ covariant spinor derivative $\Del$, and
rewrites the $N=2$ integration measure in terms of the $N=1$ measure and
explicit derivatives $Q\equiv\Del_2$ whose action on various fields is known:
$\int d^4x\, d^8\q\to \int d^4x\, d^4\q \, Q^2\qb^2$.

The leading term in a momentum expansion of the $N=2$ superspace is the
imaginary part of a ($N=2$) chiral integral of a holomorphic function $\FF(W)$
\cite{calf}
\be{calf}
S_\FF={\ci}m\int d^4x\, d^4\q \FF(W)\ .
\eq
It has the familiar $N=1$ expansion
\be{n1calf}
S_{\FF}&=&{\ci}m \int d^4x \,d^2\q\,\left[
\shalf \FF_{AB}(Q^\a W^A)(Q_\a W^B) +\FF_A (Q^2 W^A)
\right]|\nonumber\\
&=&{\ci}m \left[ \int d^4x \,d^2 \q\, \shalf\FF_{AB}(\f)\,W^{A\a} W^B_\a
+\int d^4x \,d^2\q \,d^2 \bar\q\,\FF_A(\f) \,\fb^A\ \right] .
\eq
The next term in a momentum expansion of the $N=2$ superspace
effective action for Yang-Mills theory is a full superspace integral of a real
function $\HH(W,\bw)$ \cite{h,BMM}:
\be{N2act}
S_\HH =\int d^4x\, d^8\q ~\HH(W,\bw)\ .
\eq
The $N=1$ superspace expansion of this is \cite{BMM}
\be{N1act}
S_\HH \!&=& \!\int d^4x \,d^2\q \,d^2 \bar{\q}\,
\left(g_{A\bar
B}\left[-\shalf\Del^{\a\ad}\f^A\Del_{\a\ad}\fb^B
+i\bar W^{B\ad}(\Del^\a{}_\ad W^A_\a
+\Gamma^A_{CD}\Del^\a{}_\ad\f^CW^D_\a)\right.\right.\ret
&&\qquad\qquad\qquad\qquad~~~
\raisebox{0.0em}[1.em][.9em]{$-(f^A_{CD}\,\bar W^{B\ad}
\f^C\Delb_\ad\fb^D  + f^B_{CD}\,W^{A\a}\fb^C\Del_\a\f^D)$}\\
&&\qquad\qquad\qquad\qquad~~~\left.
+(\Del^2\f^B+\shalf\G^{\bar B}_{\bar C\bar D}\bar W^{C\ad}\bar
W^D_\ad) (\Delb^2\fb^A+\shalf\G^A_{EF}W^{E\a}W^F_\a)\right]
\nonumber\\ &&\qquad\qquad\qquad~~\left.
+\squart R_{A\bar BC\bar D}(W^{A\a}W^C_\a\bar W^{B\ad}\bar W^D_\ad)
-i\mu_A(\shalf\Del^\a W^A_\alpha
-f^A_{BC}\f^B\fb^C) \right)\,\, ,  \nonumber
\eq
where $g,\G,R$ are defined in terms of the partial derivatives of $\HH$
\be{geodefs}
g_{A\bar B}=\HH_{A\bar B}\ ,\qquad\Gamma^A_{BC}=g^{A\bar
D}\HH_{BC\bar D}\ ,\qquad R_{A\bar BC\bar D}=\HH_{AC\bar B\bar
D}-g_{E\bar F}\Gamma^E_{AC}\Gamma^{\bar F}_{\bar B\bar D}\ ,
\eq
and $\mu$ is the moment map defined by:
\be{mom}
f^C_{AB}\HH_CW^B=\eta_A(W)+i\mu_A(W,\bar W)\ ,
\eq
for $\eta$ an arbitrary holomorphic function of $W$.

\section{Possible corrections}

We now look for possible corrections to masses and kinetic terms of various
fields.  By supersymmetry, it is sufficient to study only the scalar fields
$\f$, but as a consistency check, we consider all the bosonic fields, and find
the effective corrections for the gauge bosons with field strengths $f_{\a\b}$
as well. To find the mass of $\f$, we must also find the corrections for the the
vector auxiliary field $\DD$. We consistently ignore higher derivative terms:
terms with more than two derivatives on $\f$ or terms with any derivatives on
$\DD$ or $f_{\a\b}$; these terms are presumed to be artifacts of expanding the
nonlocal effective action in powers of derivatives, and not to contribute to
masses of physical states.\footnote{They may contribute to the masses of
nonpertubative states such as monopoles \cite{cru}.}

The last term in (\ref{N1act}) is an $N=1$ K\"ahler potential, and can give rise
to corrections to the kinetic terms of the scalars $\f$, the terms linear in
$\DD$, and the mass terms of the gauge bosons. Corrections to terms quadratic
in $\DD$ and $f_{\a\b}$ arise from a variety of other terms in the effective
action, and it is gratifying to see them add up to give a consistent result.

In \cite{BMM}, we argued that on dimensional grounds and by
anomalous $R$-invariance, $\HH$ has the form
\be{hhhom}
\HH=\HH^0+c\left(\ln\frac{\f^2}{\L^2}+g^0(\f)\right)
\left(\ln\frac{\fb^2}{\L^2}+\bar g^0(\fb)\right)\ ,
\eq
where $\HH^0(\f,\fb)$, the holomorphic function $g^0(\f)$, and its
conjugate $\bar g^0(\fb)$ are all functions of dimensionless ratios of $\f$
and $\fb$. For the specific case of $SU(2)$, gauge invariance and $R$-symmetry
(which follows from the assumption that $\FF$ saturates the anomaly for $\f\to
e^{i\a}\f$) imply that we can take $g(\f)=0$ and $\HH^0=\HH^0(t)$ where
\be{t}
t\equiv\frac{\f\cdot\fb}{\sqrt{\f^2\fb^2}}\ .
\eq
Before proceeding with our calculation, we make a crucial observation:
whenever $[\f,\fb]=0$, $t=1$. In particular, this means that for {\em all}
points in the moduli space vacua, $\val{t}=1$. We do not see any reason
why similar properties should hold for higher rank groups.\footnote{The
function $\HH$ was introduced in \cite{h} for the massless (abelian) sector;
for $SU(2)$, the previous arguments imply that it reduces to $c\ln
(\f^2/\L^2)\ln(\fb^2/\L^2)$.}

We now turn to the actual calculation. We make a component expansion of
(\ref{N1act}) and expand it to second order in fluctuations about the vacuum:
$\f=a{\bf e}+\vf$, $\fb=\bar a{\bf e}+\bar\vf$ where ${\bf e}\equiv e^A T_A$
is a unit vector along the direction of $\val\f$: $\val{\f_A}=a e_A$.
As explained above, we drop higher derivative terms. We find
$\DD\propto[\f,\fb]$; consequently, $\DD$ is at least linear in $\vf$. We also
use $\Del_{\a\ad}\f=\Del_{\a\ad}\vf-ia[V_{\a\ad},{\bf e}]$, as well as the
following  expectation values of the derivatives of $t$:
\be{dt}
\val{t}=1\ ,\ \ \val{t_A}=0\ ,\ \ \val{t_{A\bar
B}}=\frac1{|a|^2}(\d_{AB}-e_Ae_B)\ .
\eq

After a lengthy calculation, as expected, we find no corrections to the $U(1)$
coupling constant $\tau$, but, for the electrically charged $N=2$ vector
multiplets, though the mass is unchanged, we do find an overall shift in the
coefficient of the kinetic term:
\be{shift}
{\ci}m\val{\frac{\FF'}a}\to{\ci}m\val{\frac{\FF'}a}-2\HH'(1)\ .
\eq

In \cite{lgr}, a direct relation between the coefficient
of the charged vector multiplet kinetic term and the BPS mass
formula was found.  Were this relation to persist, then the shift in
(\ref{shift}) would pose great problems, as it would
spoil duality; however, preliminary calculations \cite{cru} indicate that it is
rather the relation found in \cite{lgr} that breaks down. Of course,
all problems would disappear if the numerical coefficient $\HH'(1)$ were
to vanish. As we shall see in section 5, $\HH'(1)$ is sensitive to
infrared divergences in the theory, and when the theory is
asymptotically free, at least to 1-loop there exists an infrared
regularization which does lead to $\HH'(1)=0$. In contrast to this
infrared sensitivity of $\HH$, $\FF$ is independent of the infrared
cutoff up to rescalings of the ultraviolet scale $\Lambda$, and seems to
be a physically sensible, scheme independent object.

For higher rank gauge groups with larger dimension moduli spaces, there exist
$R$-invariant dimensionless ratios of moduli, and we expect a new
complication to arise: the shift analogous to (\ref{shift}) may not be by a
numerical (if scheme dependent) constant, but may vary over the moduli space.
In that case, it seems unlikely that even the leading effects of $\HH$ could
be tuned to vanish.

\section{Higher dimension terms and ambiguities}

Is $\HH$ unique? Are there ambiguities in it that we have not considered?
More precisely, we want to know if there are other higher dimension terms that
can contribute to the effective K\"ahler potential.

We first find a rather trivial ambiguity in $\HH$ that we can resolve
immediately: it seems possible to absorb a term proportional to the classical
action in $\HH$, that is, a contribution to the effective
K\"ahler potential $\D K=\g \f\cdot\fb$ for a constant $\g$. For gauge group
$SU(2)$, the differential equation that we solve to find $\HH(t)$ from the
effective K\"ahler potential (after subtracting out the $\FF$ contribution is:
\be{dH}
\frac{\D K}{\f\cdot\fb} =\HH'(t) \frac{1-t^2}t\ .
\eq
For $\D K= 2\g\f\cdot\fb$, we appear to find a
solution $\HH=-\g\ln(1-t^2)$; however, this is singular whenever $t=1$, that is,
in the vacuum, and therefore the coefficient of this contribution is
unambiguously set to zero (this divergence has a different form than the
infrared divergences that we discuss in the next section, and cannot be
cancelled against them).\footnote{In
\cite{BMM}, we set the coefficient to zero without analyzing it.} Indeed,
since $\HH$ cannot contribute to the K\"ahler potential
for the massless fields, and $\f\cdot\fb$ does contribute, we know without any
calculation that this term must be singular and cannot be included.

A true ambiguity comes from even higher dimension terms in $N=2$
superspace whose $N=1$ component expansions give rise to corrections to the
effective K\"ahler potential.  These can be classified, and are finite in
number.
The key observations are: (1) The expansion of the measure
$\int d^4x\, d^8\q\to\int d^4x\, d^4\q \, Q^2\qb^2$ gives a factor of $Q^2\bar
Q^2$ that must somehow be ``absorbed''. This can happen because (2) Eq.\
(\ref{cons}) implies $\{\Del_\a,Q_\b\}=iC_{\a\b}\bar W$ and (3) Eq.\
(\ref{relb}) implies $Q^2W=\Delb^2\bar W$. Thus, terms that can contribute to
the effective K\"ahler potential must contain
$1,Q\Del,\Del^2\Delb^2$, where this represents any possible
combination of these (in particular, the last includes possible
$\Del_{\a\ad}\Del^{\a\ad}$ terms).  All other terms are related to these by
complex conjugation or integration by parts, or do not contribute to the
effective K\"ahler potential. After integration by parts,
the possible terms and the contributions that they give rise to are:

\be{hdc}
\HH\too -\HH_{\bar A}f^A_{BC}f^C_{DE}\fb^B\fb^D\f^E\ret\ret
\HH^1_{AB}\Del^\a W^AQ_\a
W^B\too-2\HH^1_{AB}f^A_{CD}f^B_{EH}f^E_{FG}\fb^C\f^D\fb^F\f^G\f^H\ret\ret
\HH^2_{AB}\Del^{\a\ad}W^A\Del_{\a\ad}W^B\too4
\HH^2_{AB}f^A_{CF}f^B_{GJ}f^C_{DE}f^G_{HI}\fb^D\f^E\f^F\fb^H\f^I\f^J\ret\ret
\HH^3_{A\bar B}\Del^{\a\ad}W^A\Del_{\a\ad}\bar W^B\too-4\HH^3_{A\bar
B}f^A_{CF}f^B_{GJ}f^C_{DE}f^G_{HI}\fb^D\f^E\f^F\f^H\fb^I\fb^J\ret\ret
\HH^4_{AB\bar C}\Del^{\a\ad}W^A\Del_\a W^B\Delb_\ad\bar W^C\too
4i\HH^4_{AB\bar
C}f^A_{DG}f^B_{HI}f^C_{JK}f^D_{EF}\fb^E\f^F\f^G\fb^H\f^I\f^J\fb^K\ret\ret
\HH^5_{AB\bar C\bar D}\Del^\a W^A\Del_\a W^B\Delb^\ad\bar
W^C\Delb_\ad\bar W^D\too4\HH^5_{AB\bar C\bar
D}f^A_{EF}f^B_{GH}f^C_{IJ}f^D_{KL}\f^E\fb^F\f^G\fb^H\f^I\fb^J\f^K\fb^L\ret
\eq
These terms can also be written in manifestly $SU(2)$ invariant notation; the
only subtlety involves the $\HH^1$ term. It arises from terms proportional to
$\Del^2_{ab}W^A\Del^{a\a}W^B\Del^b_\a W^C$, and similar terms related by
partial integration.

We have not analyzed the effects of these higher dimension
terms in detail; though they may lead to modifications of the explicit 1-loop
$\HH^0$ found in \cite{BMM}, we believe that they do not change any of our
qualitative conclusions. In particular, it is clear that they can only
contribute to the massive sector, as they all involve commutators.

\section{Infrared Issues}
In all the previous sections, we have implicitly assumed that $\HH$ is
well-defined. However, if we use the explicit 1-loop $\HH$ found in \cite{BMM},
or equivalently, look directly at the 1-loop effective K\"ahler potential,
the contribution of the vector multiplet to the coefficient of the kinetic
terms of the massive fields ($\HH'(1)$) is divergent.  Examining the explicit
calculation \cite{BMM,MMR}, we have a contribution
\be{varK}
\D K\propto \int^{\L^2}dk^2\, Tr[\ln(k^2+M)-\ln(k^2)]\ ,
\eq
where $M$ is the matrix $\{\f,\fb\}$; for $SU(2)$, $M$ has eigenvalues
$s,\frac12(s\pm u)$, where $s=\f\cdot\fb$ and $u=\sqrt{\f^2\fb^2}$. The last of
these, $\frac12(s-u)$, vanishes in the vacuum at all points in moduli space.
Thus we see that the problem is an infrared divergence. This leads us to try to
define a Wilsonian effective action.

The naive procedure for doing this is to introduce an infrared cutoff $\mu^2$.
However, as discussed in \cite{BMM}, such a cutoff is incompatible with special
geometry: restricting to the abelian sector $[\f,\fb]=0$, where $\HH$ can give
only higher derivative contributions \cite{h} and cannot contribute to the
effective K\"ahler potential, we find a change $s\ln
s\to(\mu^2+s)\ln(\mu^2+s)$. This {\em cannot} be written in the form
${\ci}m(\fb\FF'(\f))$ for any holomorphic function $\FF(\f)$, and is thus
incompatible with $N=2$ supersymmetry.

We therefore propose a new Wilsonian effective action with a {\em
field-dependent} cutoff (somewhat in the spirit of \cite{ds}).  Since the
mass-scale in the theory is set by $\left< s\right>$, we introduce an infrared
cutoff $\xi s=\xi\f\cdot\fb$ where $\xi$ is a dimensionless constant. Though
the physical interpretation of this cutoff is somewhat unclear (we are cutting
off our integration on the scale of the classical effective field $\f$), to
this order in the momentum expansion the definition is unambiguous, and to
higher orders there is ``ordering'' ambiguity.

With this definition, we have calculated
the one loop contribution to $\HH$. We subtract a
contribution proportional to $s\ln(u/\L^2)$, as this goes into $\FF$; this
is $\xi$-independent aside from a $\xi$-dependent
redefinition of $\L$ which cancels terms proportional to the classical
action $s$ as discussed at the beginning of the previous section. The remaining
contribution to the effective action coming from integrating over the vector
multiplet is
\be{vector}
\D K_V=-\frac{s}{\left(4\p\right)^2}\left\{\left(2\xi-1\right)\ln\left(t\right)
  +\xi\ln\left(\xi\right)+\left(\xi+1\right)\ln\left(\xi+1\right)
  -\half\left(2\xi+1\right)\ln\frac{\left(2\xi+1\right)^2t^2-1}{4}
  \right.\ret \left.
  -\frac{1}{2t}\ln\frac{\left(2\xi+1\right)t+1}{\left(2\xi+1\right)t-1}
 \right\}\ .\ret
\eq
The $\xi$-dependent classical term that we have absorbed by rescaling $\L$ is
\be{vconst}
-\frac{2s}{(4\pi)^2}(1+\xi\ln\xi-(\xi+1)\ln(\xi+1))\ .
\eq
Matter in the adjoint representation contributes
\be{adjmatter}
\D K_{adj}= -\frac{2s}{\left(4\p\right)^2}\ln\left(t\right)\ .
\eq
Note that in this case, the classical term
\be{vadj}
\frac{2s}{(4\pi)^2}(1+\xi\ln\xi-(\xi+1)\ln(\xi+1))
\eq
has {\em all} the $\xi$-dependence. Finally, matter in the fundamental
representation contributes
\be{fundmatter}
\D K_{fund}= \frac{s}{\left(4\p\right)^2}\left\{2\xi\ln\left(t\right)
  -\frac{1}{4}\left(4\xi+1\right)
   \ln\frac{\left(\left(4\xi+1\right)^2-1\right)t^2+1}{\left(4\xi+1\right)^2}
  \right.\nonumber\\ \left.
  -\frac{1}{4t}\sqrt{t^2-1}
   \ln\frac{\left(4\xi+1\right)t+\sqrt{t^2-1}}
     {\left(4\xi+1\right)t-\sqrt{t^2-1}}
 \right\}\ .
\eq
The classical term that has been absorbed is
\be{vfund}
\left(\frac14\right)\frac{2s}{(4\pi)^2}
(1+4\xi\ln4\xi-(4\xi+1)\ln(4\xi+1)+\ln4)\ .
\eq

The three contributions (\ref{vector},\ref{adjmatter},\ref{fundmatter}) should
be compared to the contribution from $\HH$, which is
\be{Hcont}
\D K= u\HH'\left(t\right)\left(1-t^2\right)\ .
\eq
To calculate the infrared regulated $\HH$, we equate these expressions, and
find a first order differential equation for $\HH$ as in \cite{BMM}.

We can eliminate $\HH'\left(1\right)$ by fine tuning the parameter
$\xi$. To see when this can be done we take the contribution to the
one-loop effective action coming from $n$ matter multiplets in the
adjoint representation and $m$ matter multiplets in the fundamental
representation. Expanding this function of $t=1+\e$ for small $\e$ and
requiring that the $\e$ dependent piece vanishes gives us an equation
for $\xi$
\be{ceqn}
 \ln\left(\frac{1+\xi}{\xi}\right)=4-4n-
   \frac{2m\left(2\xi+1\right)}{\left(4\xi+1\right)}
\eq
This equation has a solution with $\xi$ real and positive only if
$m<4,n=0$, that
is, if the theory is asymtotically free; negative $\xi$ doesn't make
physical sense for an infrared cutoff. For the conformally invariant cases
$m=4,n=0$ and $m=0,n=1$, (\ref{ceqn}) is solved by $\xi=\infty$;
physically, this corresponds to completely suppressing all quantum
corrections.  Since that correctly gives $\D\FF=0$, this is
expected.

\vskip 1em
\noindent{\large\bf Acknowledgements}
\vskip 1em

\noindent{It is a pleasure to thank Jan de Boer, Kostas Skenderis, Gordon
Chalmers, Marc Grisaru, Erick Weinberg, Mike Dine, and Yuri Shirman for
stimulating conversations. The work of MR and FG-R was
supported in part by NSF grant No.~PHY 9309888. The work of UL was supported in
part by NFR grant No.~F-AA/FU 04038-312 and by NorfA grant No.~94.35.049-O.}

\end{document}